\documentclass[manuscript, sigconf]{acmart}

\hypersetup{
     colorlinks=true,
     linkcolor=blue,
     filecolor=blue,
     citecolor = black,      
     urlcolor=cyan,
     }
     
\usepackage{booktabs}  
\usepackage{verbatim}  
\usepackage{balance}
\usepackage{float}

\usepackage{array}
\newcolumntype{L}[1]{>{\raggedright\let\newline\\\arraybackslash\hspace{0pt}}m{#1}}
\newcolumntype{C}[1]{>{\centering\let\newline\\\arraybackslash\hspace{0pt}}m{#1}}
\newcolumntype{R}[1]{>{\raggedleft\let\newline\\\arraybackslash\hspace{0pt}}m{#1}}
\usepackage{multirow}

\usepackage{listings} % code
\setcopyright{rightsretained}
\acmISBN{978-1-4503-5628-2/18/08}

\copyrightyear{2026} 
\acmYear{2026} 
\acmConference[SIGCSE '26]{Proceedings of the 56th ACM Technical Symposium on Computer Science Education}{Feb 18--21, 2026}{St. Louis, MO, USA}

\acmPrice{15.00}

\begin{document}
\title{A “watch your replay videos” reflection assignment on comparing programming without versus with generative AI: learning about programming, critical AI use and limitations, and reflection}
%A reflective community of practice approach for teaching technology ethics and professional skills: r

\author{Sarah ``Magz'' Fernandez}
\email{ sarah.fernandez@maine.edu}
\affiliation{%
  \institution{University of Maine}
   \city{Orono}
  \state{Maine}
  \country{USA}
}

\author{Greg L Nelson}
\email{gregory.nelson@maine.edu}
\affiliation{%
  \institution{University of Maine}
   \city{Orono}
  \state{Maine}
  \country{USA}
}

\begin{comment}
    \author{Greg L Nelson, Sarah ``Magz'' Fernandez}
\email{gregory.nelson@maine.edu, sarah.fernandez@maine.edu}
\affiliation{%
  \institution{University of Maine}
   \city{Orono}
  \state{Maine}
  \country{USA}
}

    \author{Sarah ``Magz'' Fernandez}
\email{ sarah.fernandez@maine.edu}
\affiliation{%
  \institution{University of Maine}
   \city{Orono}
  \state{Maine}
  \country{USA}
}
\end{comment}

\renewcommand{\shortauthors}{}
\renewcommand{\shorttitle}{Comparative watch the replay reflection}

\begin{abstract}
Generative AI is disrupting computing education. Most interventions focus on teaching
GenAI use rather than helping students understand how AI changes their 
programming process. We designed and deployed a novel comparative video reflection assignment adapting the
Describe, Examine, then Articulate Learning (DEAL) framework. In an introductory software engineering course,
students recorded themselves programming during their team project two times: first
without, then with using generative AI. Students then analyzed their own videos using a
scaffolded set of reflection questions, including on their programming process and human,
internet, and AI help-seeking. We conducted a qualitative thematic analysis of the
reflections, finding students developed insights about planning, debugging,
and help-seeking behaviors that transcended AI use. Students reported learning to slow
down and understand before writing or generating code, recognized patterns in their
problem-solving approaches, and articulated specific process improvements. Students also
learned and reflected on AI limits and downsides, and strategies to use AI more critically, including
better prompting but also to benefit their learning instead of just completing tasks.
Unexpectedly, the comparative reflection also scaffolded reflection on programming not involving AI use, and even led to students spontaneously
setting future goals to adopt video and other regular reflection. This work
demonstrates structured reflection on  programming session videos can develop
metacognitive skills essential for programming with and without generative AI and also
lifelong learning in our evolving field.

\end{abstract}



\maketitle

\section{Introduction}
 
Learning to reflect is a part of being a reflective practitioner and a professional computer scientist, because reflection is an essential professional skill for improving over time.
Reflection and iterative improvement are also a part of professional skills and widespread practices, such as in agile software development during sprint retrospectives, which are a kind of reflection.

Generative AI is transforming how we program; this ongoing situation and need for learning makes reflection even more important, but our field knows relatively little about how to foster reflection on critical and ethical generative AI use specifically \cite{fengAutomationCognitionRedefining2024, pratherHypeComprehensiveReview2025, pratherWideningGapBenefits2024a}.

The education research literature has found that careful scaffolding of reflection \cite{shaoEffectsRegulatedLearning2023, theobaldSelfregulatedLearningTraining2021, zuoEffectsUsingScaffolding2023}
greatly increases the quality of reflection and learning outcomes. The most common scaffolding is via a sequence of reflection steps, such as questions, which is called a script \cite{wangAdaptableScriptingFoster2017}
in the literature (not the same as a computer program script). For example, the DEAL reflection framework involves three sequential steps: Describe, Examine, and Articulate Learning \cite{ashGeneratingDeepeningDocumenting2009}. This framework is general and applicable to multiple domains, but must be implemented to fit the domain and subject of reflection. Scaffolding reflection includes designing the experience and emotional aspect, such as in usability testing, where students share their app and watch users try to use it, but without speaking or being able to do anything (which can be painful but very informative). Students observe directly and in detail where users struggle, spin their wheels, and misunderstand things, and they learn from it.

Drawing on those, we used and extended the DEAL framework to design a comparative video reflection activity, for doing a programming session without generative AI, then with generative AI. For the Describe step, we had students record and then later review a screen recording of each programming session. Next, for the Examine step, students reflected on challenges, progress, programming sub-activities such as planning, help and information seeking, writing, and debugging, then the differences between the sessions and potential causes of those differences. This general design idea of watching and comparing videos for reflection appears novel in computing education. For the Articulate Learning step, students synthesized their analyses, what they learned from that, how it was learned, why that learning matters, and what future goals they might set based on their learning.

We deployed this assignment in an undergraduate introductory software engineering class in Fall 2024, and answered the following research questions using a qualitative thematic analysis of student written submissions for the reflection assignment: 

\begin{enumerate}
    \item \textbf{RQ1} When students reflect on their programming using videos and DEAL, what do they report learning related to a) programming while using generative AI tools, and b) programming in general?
    \item \textbf{RQ2} When students reflect on their programming using videos and DEAL, what ideas for acting on their learning do they report in their reflection?
  
\end{enumerate}

\section{Related Work}

There is good work on students reflecting on their programming, finding that it is generally positive; however, we have found no research on computing students reflecting using an assignment following the DEAL framework (a particular research-based reflection design framework). The DEAL reflection framework originated in the context of domain general research on service learning (such as doing a class project for a non-profit) \cite{ashGeneratingDeepeningDocumenting2009, robledoyamamotoCISingServiceLearning2023}. Within computing education, there is work on computing students reflecting on service learning with essays and journals over time, but not specifically scoped to their programming process or a couple programming sessions \cite{robledoyamamotoCISingServiceLearning2023,christianssonTeachingParticipatoryDesign2018}
Beyond that, many instructors do include some reflection in their courses but much of it is not published.

Much of the work in computing education focuses on reflection about comparing, evaluating, or explaining solutions to problems, rather than on reflecting on a problem solving process or programming session itself. This is often because the reflection is designed to accomplish worthy technical knowledge or narrow skill learning objectives; for example, for the learning objective “utilize the suitable map and reduce operations in Spark to transform and aggregate data” the reflection “Think of the solution you were attempting. Is this the same as what the bot presented in the chat? Take this time to explain the solution to each other” where the bot presented a solution that met the learning objectives \cite{sankaranarayananCollaborativeProgrammingWorkRelevant2022,sankaranarayananCollaborativeReflectionFlow}. While that system involved a conversational agent to give hints and scaffold reflection of that kind, students did not reflect on their overall programming process, experience, or on using the agent itself \cite{sankaranarayananCollaborativeReflectionFlow, naikGeneratingSituatedReflection2024, sankaranarayananDesigningLearningCollaborative2020}. 
Similarly, some work includes collaborative reflection and students comparing, extending, and challenging other students reasoning \cite{adamsonAgileApproachAdapting2014, sankaranarayananCollaborativeReflectionFlow} or pedagogical code review finding issues with code itself, not the process of coding \cite{hundhausenTalkingCodeIntegrating2013}. Other work tries to scaffold metacognitive reflection at a more tactical level such as while solving a particular kind of problem \cite{kazemitabaarExploringDesignSpace2025, fengAutomationCognitionRedefining2024}. 

While some excellent work closer to ours has students reflect on their GenAI use \cite{grandeStudentPerspectivesUsing2024, lyuWillYourNext2025}, only a few works have reflection on problem-solving process and GenAI use. But we are unaware of prior published work on assignments for students reflecting to compare their process when not using GenAI versus when using GenAI, nor comparing sessions. 
Some work has students reflect not on the programming process but instead comparing solutions, such as comparing one's code vs. code solely done by generative AI \cite{fengAutomationCognitionRedefining2024} e.g. “Compare the solutions and comment on the similarities and differences. Submit both solutions and your critique of the main differences between them.” 
The closest work was research on understanding students; it did a brief content analysis of student responses to an optional reflection question after an assignment on describing their process to complete the assignment, when and why they used AI, and how it was helpful or not \cite{margulieuxSelfRegulationSelfEfficacyFear2024}. Lyu et al analyzed some student open-ended reflections, and their course included using GenAI, pairing with a human, or both, but there is no description of the reflection questions or topics \cite{lyuWillYourNext2025}.
Long before GenAI but also in response to GenAI, work has called for increasing integration of metacognition in computing, including “critical thinking and reflection” \cite{fengAutomationCognitionRedefining2024,tankelevitchMetacognitiveDemandsOpportunities2024a,cakirogluModelDevelopActivities2023,loksaMetacognitionSelfRegulationProgramming2022a} .

While video reviewing is common in qualitative research and for coaching \cite{fukkinkVideoFeedbackEducation2011, leungAdoptionVideoAnnotation2021}, having students review videos of their work as an assignment is rare in computing education, and comparing sessions appears novel. Students have recorded retrospective video explanations of their solutions, such as on assignments \cite{vandegriftGivingVoiceProblemSolving2024} or as "post-exam videos" \cite{vandegriftPostExamVideosAssessment2022}, like an oral exam where they explain and correct mistakes as they explain their solutions. Some works study video reflection for teacher training \cite{broneakExploringUseVideo2019,allisonClassifyingCharacteristicsEffective2023}.

There is work on reflecting on code editor replays (keystroke level) on small programming problems, but not videos which also show more information seeking and other behaviors on the screen. We are aware of only two works. In 2021 Ditton et al piloted editor replays finding some benefits and student frustrations around value of reviewing them \cite{dittonExternalImageryComputer2021}, and in 2023 Xie et al found students who reflected on code replays exhibited better self-regulation like planning more often \cite{xieDevelopingNoviceProgrammers2023}. 
There are also research studies where the researchers  analyze student programming sessions while using generative AI with video \cite{pratherWideningGapBenefits2024a}, which involves students thinking out loud or reflecting in a diary or survey question \cite{margulieuxSelfRegulationSelfEfficacyFear2024}, 
but not as an assignment.

\section{Design of reflection assignment} 
We deployed this reflection assignment in an undergraduate software engineering class, delivered synchronous fully online, in Fall 2024 at [Anonymized], a northeastern rural university.  For context, generative AI use was covered explicitly in this course, including basic prompt engineering with roles and output formatting, practice using GenAI for feedback on requirements, a class session where a TA demo'd using Cursor on their personal website project, and a class session on generating tests using GenAI. The students recorded programming sessions in the context of working on a software engineering team project, with 7 of 8 projects involving web development using React and Firebase (one was a Discord bot). The screen recorded videos of their programming sessions were recorded by students before the assignment was released, though they were told they would be used for a “watch the replay” reflection assignment. Students were reminded to, when recording, to only record their screen, turn off their notifications, and edit out of their videos anything they might feel uncomfortable sharing. 

We applied the DEAL framework, which is a high level architectural pattern for designing reflection. The sequential steps in DEAL are Describe, Examine, and Articulate Learning \cite{ashGeneratingDeepeningDocumenting2009}. Normally each step has a set of reflection questions to respond to, usually in writing. We detail our reflection assignment below and also have published the full assignment with a Creative Commons license \href{http:\\anonymized.com}{here}.

In our design for DEAL reflection, the Describe step was to a) work on their team software project while recording a video of a session without using GenAI, then a session with using GenAI, then b) later review each video by watching it. The instructions were to record a programming session while working on their class project (programming they would do in the course anyways) during their second to last project milestone/sprint, then record another one while working on their final milestone/sprint. In contrast, without videos, traditionally in DEAL reflection this step involves writing long descriptions of what happened, focusing on facts and observations \cite{ashGeneratingDeepeningDocumenting2009}.

In the Examine step, students answered a set of written reflection questions in the following order. To answer these, the assignment had students fill in a table with four columns with headers "Reflection question", "Answer for Session 1", "Answer for Session 2", "Answer for Compare the differences". Each row had one reflection question, for example, "What were you working on?". First, students were instructed to watch video 1 then answer all of the Examine questions, then repeat that for video 2 (we will describe the "compare the differences" column in the next paragraph). The sequence of questions started with questions to make sense of what happened at a higher level: 1) What were you working on? 2) What progress was made? 3) If you did pair programming, who did you pair with?, 4) What resources, people, or tools did you work with? 
5) What challenges did you encounter?, and 6) How did you respond to each challenge?.
Next, students answered detailed questions on roughly what percent of their time was spent in different categories while programming, related to basic steps in the programming process. The first three time categories were: planning, writing or editing code manually, debugging or understanding code. Next were three time categories related to information seeking and human collaboration: looking up and using examples, looking up and using API documentation, asking another person for help, and working with another person. Finally, there were time categories for using generative AI: using generative AI to write code, using generative AI to understand code or ask questions about APIs,etc , using generative AI to debug code, using generative AI for planning. 

In the Examine step, students also compared their two programming sessions, without generative AI, and with generative AI. First, they answered what could be improved about each session, and what they did differently for the second session (usually just using generative AI). For each reflection question from the prior steps, students answered “Compare and Evaluate: write about 3-5 bullet points total per question on a) the differences between the two videos / programming sessions, b) what caused those differences.” Those answers were written in the last column of the reflection questions table, each in a cell with "differences:", some whitespace, then "causes of differences:". 

Next, for the Articulate Learning step, students answered the following questions. For context, video 1 was the session without using generative AI, and video 2 was the session with using generative AI. The last five questions are from the DEAL framework examples in \cite{ashGeneratingDeepeningDocumenting2009}, with the last question broadened:
\begin{enumerate}
    \item  Summarize the most significant differences between your software development session in Video 1 vs. Video 2
    \item Summarize the causes of those differences and what their effects/consequences were
    \item What did I learn?
    \item How, specifically, did I learn it?
    \item Why does this learning matter, why is it important?
    \item In what ways will I use this learning?
    \item What goals might I set in accordance with what I have learned in order to improve myself and/or the quality of my learning and/or the quality of my future?
\end{enumerate}

The assignment instructions also included instructions to scope the reflection and help students manage their time, with time guidelines for each video, and explicit length norms for each question, to make the assignment manageable for students. For nearly all of the questions, the instructions were to “Write about 3-5 bullet points per question” (except the two articulate learning questions on summarizing the most significant differences between the videos, and the causes and effects). When watching the video, we instructed students to be efficient i.e. "watch each video at 2x or higher speed, click through the video, etc."

\section{Methods} \label{method}

Thirty nine students enrolled in the course, with 36 active in the course in the last three weeks of the semester. No students opted out of the research. This research was reviewed by our institutional ethics board and approved.  Twenty eight (93\%) of 30 responded to a course demographics survey. Of the 28 respondents, 19 (68\%) identified as men, 8 (29\%) as women, and one (4\%) as non-binary. They included 14 (50\%) 3rd year students, 8 (29\%) 4th year, and 2 (7\%) 2nd Year students, one (4\%) 6th year, and 3 (14\%) graduate students. Twelve (42\%) reported a disability or mental health condition, with 9 (32\%) Attention Deficit, 5 (18\%) a "Mental health condition", 2 (7\%) a Learning disability, 1 (4\%) Autism, 1 (4\%) Deaf or hard of hearing, and 1 (4\%) PTSD. Beyond that, three others responded unsure or preferred not to answer.
Twenty (71\%) identified as White or Caucasian, 2 (7\%) Asian, two (7\%) Black or African American, 1 (4\%) Hispanic or Latinx, and one (4\%) preferred not to disclose. Ages mostly ranged from 19 to 23 years (21 respondents), 
with two (7\%) 24-year-olds, one (4\%) 25, and four (14\%) between 28 and 33.

At our institution, prior to this course, students generally had not done group projects, and pair programming is not taught in the curriculum. Four out of thirty students chose to do their second session with GenAI and human pair-programming. While a packet on pair programming was distributed, students did not receive explicit instruction on pair programming itself, but did experience virtual pair programming and setup VSCode for that in the 75 minute class exercise on test-driven development, including using GenAI for generating tests.

For our analysis, we conducted a qualitative thematic analysis \cite{braunUsingThematicAnalysis2006,hammerConfusingClaimsData2014} of the student’s written reflection submissions. Our analysis team was a person unaffiliated with the university and the course instructor.
We first wrote memos on student whole answers to each question, reading each student as a whole, which we then discussed. We then decided to, based on the bullet point format used by students, to scope our analysis to specific questions to make a large affinity diagram of all of the responses, clustering them to then create themes. We chose questions from the “articulate learning step” of their reflections for our analysis: \textit{“What did I learn?”} and \textit{“In what ways will I use this learning?” }, then
\textit{“What goals might I set in accordance with what I have learned in order to improve myself and/or the quality of my learning and/or the quality of my future?”}. We chose these questions to balance a manageable set of data to analyze while also still covering what students learned, as well as
how students may use their learning and the future goals they might set, which helps gauge supporting their reflection to be actionable for putting that learning into practice. 
As a reminder, these questions were answered by students without any example reflection question answers.

Throughout our analysis, rather than calculating inter-rater reliability, we prioritized resolving disagreements through collaborative discussion to deepen shared understanding, consistent with practices for qualitative rigor in computing education \cite{salacFundsKnowledgeUsed2023, eversonKeyReducingInequities2022,hammerConfusingClaimsData2014}. We came to agreement on and chose to report themes to cover important patterns, including any potential harms, in the data.
Each theme had at least 3 unique students.

\section{Results} \label{results}

As noted in our method, we are reporting \textbf{themes}, not asserting specific numbers of students experienced them, but instead broadly reporting the diverse kinds of learning and goals set by students.

\subsection{RQ1a: Learning related to programming with generative AI}

\textbf{AI Downsides and Negative Impacts on Learning:}
Students reported learning about downsides and negative impacts from using GenAI. For example, P32: "I also learned that using Generative AI causes me to learn less than actually coding manually.", and P10: "I learned that ai can be helpful when used in the right way, but it can also be unreliable, and I can see how it could have negative effects on my learning". Other students learned that over-reliance can lead to problems, like P20: "Relying too heavily on generative AI leads to different problems.", and other students learned AI use can be situationally counter-productive, such as P11: "Generative AI can sometimes be extremely counterproductive, especially when used on an already strange piece of code."
Another student, P14, commented on how writing code from scratch themselves can be more productive than using code from GenAI or the internet.

\textbf{AI Limitations:}
Students reported learning about limitations of the GenAI tools. For example, P18 : "I learned that AI for code generation has flaws.", P13 : "Sometimes Gemini’s summaries and reading comprehension suck.",
and P26 : "Github Co-Pilot isn’t as good at creating tests." This also included reinforcing prior learning, like P19: "I reinforced the understanding I already have that when not used carefully AI often creates a lot of problems with junk code or in the case of the first video doesnt know enough context and will lead you down rabbit holes for simple solutions.", 

\textbf{AI for Debugging and Critical Use of AI for Debugging:} Students reported learning around debugging skills. This included that AI was useful for debugging, such as P21: ``...[AI] can explain the issues well too.", but also included critical use and non-use of AI for debugging, for example, P1: "How to use AI more effectively in debugging", P8: "Manual debugging and understanding code is key whether using AI or not.", and P11: "I continue to be best at fixing bugs when I do so by reading examples or documentation."

\textbf{AI Prompt Engineering:}
Students reported learning skills around prompt engineering. For example, student reflections included P8: "Prompting AI effectively is important to get the results you want.", P20: "Effective communication with AI tools is critical to getting useful outputs.", and
P14 : "How [AI] is a helpful resource but can’t fully help you write code due to it missing some context."

\textbf{AI Utility and Benefits:}
Students reported learning about the utility and benefits of Generative AI in programming tasks. For example, student reflections included P21: "AI is actually very good at specific coding problems, and it can explain the issues well too.", P4: "I learned that if used correctly GenAi can be a useful tool.", and P30: "AI tools are really powerful and can help me streamline certain processes like writing boilerplate code and comments."

\textbf{AI Increases Productivity:}
Students specifically reflected that Generative AI increased their productivity and efficiency. For example, P15: "Generative AI can significantly enhance productivity when used effectively.", P3: "...GenAI can save time.", and P8: "GenAI is super useful for debugging and faster coding."

\subsection{RQ1b: Learning related to programming in general}

\textbf{Need for Background and Understanding}
Students reported learning the importance of having a solid background and understanding of tools, APIs, and their project codebases before implementation. For example, student reflections included P30: "I had no chance of coding manually in the first video, because I lacked the understanding", P8: "Manual debugging and understanding code is key whether using AI or not.", and P1: "Researching libraries’ feasibility ahead of time before implementing them."

\textbf{Planning:} Students reported learning the value of planning. For example, student reflections included P27: "I learned that planning ahead of time is incredibly important...", P20: "Planning and structuring work before starting can help reduce issues later.", and P30: "I learned that having a plan on what to do really helps. When I tried random code bits aimlessly, I lost a lot of time."

\textbf{Human help seeking:} Students reported learning to seek help from people. For example, P27: "...it is sometimes way more efficient to just ask for help.", P17: "I learned how to pair program and ask for help and be vulnerable when I was confused and needed help from others." At least one student learned that human help seeking was preferable to AI, for example, P32: "I learned that asking a person sometimes is better than asking AI."  

\textbf{Metacognitive insights about learning:} Students reported positive metacognitive insights about how they learn programming. For example, student reflections included  P30: "It’s okay to start a feature even when you don’t know what to do. You try a few things, and start building your knowledge base. And you try to learn each time you fail.", P5: "If a task is taking a while a going nowhere, move on to something else.",  and P13: "Learning the API is important, but this can’t be done (easily) without coding."

\textbf{Self-learning and professional identity}
Students reflected learning more about themselves and developing professional identity. For example, P21: "I won’t do test driven development (At least not that way). I’m a game dev, it’s called playtesting",  P10: "I learned about myself as a programmer", and P13: "This learning will be used in a never-ending struggle for self-learning and improvement."

\textbf{Specific technical learning}
Students identified specific skills they learned in addition to GenAI. Examples include student reflections such as P17: "I learned how to use css to get my UI to look a specific way.", P29: "Learned how to use .md files.", and P16: "How the server receives request."

\subsection{RQ2: Student-written goals for acting on their learning from their reflection}
Students in the reflection generated very diverse goals for acting on their learning, including more skillful and critical GenAI use, but just as often on improving programming generally with better help-seeking, human connection and collaboration, and even purposefully not coding with GenAI. 

\textbf{Balance in using GenAI:}
Students set the goal of achieving a balance between AI use and other skills. For example, P8: "Balance using AI with traditional tools so i keep my core programming skills while also working faster.", P20: "Develop better planning and organizational skills to reduce reliance on AI and reactive problem-solving.", and P1: "Maintain a balance between manual problem-solving and using AI to debug problems."

\textbf{Better Prompt Engineering:}
Students set the goal to learn better prompt engineering. Examples include P20: "Set a goal to allocate time for improving communication with AI tools, such as learning advanced prompting techniques.", P15: "Focus on learning more about efficient prompt engineering and API usage to streamline future projects.", P24: "I will set a goal to improve my AI prompting and documentation research."

\textbf{More critical use of AI:}
Students set the goal being more critical in their use of AI. For example, P8: "Build a habit of reading through AI generated code to understand the logic and continue learning my skills past school.", and P20: "Enhance my ability to debug and refine AI code by deepening my understanding of the underlying concepts.". This also included hesitant students opening up to more but critical AI use, for example, P4: "I want to set the goal of being more open minded when it comes to certain uses of AI. I generally dislike it, especially when it comes to anything creative. But using it to direct some code, or give a suggestion on how to refactor something is useful, and doesn’t impact my creativity or harm others."

\textbf{Try out using AI more:}
Students set general goals of just trying to use AI more. For example, P32: "I will be more open to using AI, instead of having a sense of guilt when using it to write programs.", P19: "Become more effective at using AI generated code in a clean and concise way that makes my coding more efficient instead of less.", P15: "Develop a consistent approach to using generative AI for debugging, understanding APIs, and optimizing code."

\textbf{Adopting a practice of regular self-reflection:}
Students set the goal of more self-reflection. Examples of self-reflection related goals include P10: "Goals I will set for myself: Keep self reflecting and checking my progress as a student and a programmer.", P18: "I’d like to go back and review my old code every once in a while to see my improvement.", and P25: "Sometimes I should try changing something a little bit in a way I approach a particular activity to eventually optimize it. The week after finals I am going to think of such activities and then for 30 days notice them and try to improve."

\textbf{Adopting a practice of regular self-reflection with videos:}
Students included goals around reflecting on video recordings of themselves. Examples included P32: "I will analyze my work in the future through video in time intervals (one month apart) to see how I have improved as a programmer.", P31: "watching myself work month to month to see what im struggling on.", and P5: "Maybe record my work again in the future."

\textbf{Collaborate with other humans:}
Students set the goal to collaborate with others. Examples include, P27: "I will set goals such as, reviewing my code with a peer or mentor, this way I can receive the help i need ahead of time rather than being stubborn and wanting to finish the project on my own.", P31: "do more pair programming to see better ways other people do things.", and P8: "Actively look for more opportunities to collaborate with others on programming."

\textbf{Ask for help sooner instead of struggling:}
Students set the goal to ask for help from others. For example, P17: "Set a time limit when I am stuck on something before I reach out for help so I don’t spend a significant amount of time trying to understand something.", P16: "ASK MORE QUESTIONS!!! Write down questions and ask them.", and P22: "don’t sit around and struggle on things by yourself ask for help sooner."

\textbf{Coding without AI:}
Students set the goal to learn coding without relying on AI or for comparative experimentation. Examples include P18: "I’d like to practice coding without AI, to guarantee that I have a strong foundation, regardless of what technology there is.", P30: "I might save the day with the code I borrowed on the internet, or the code that AI wrote but there will be a day when I need to expand on my project...It is not a good habit to write code that I don’t understand." This also included self-experimenting, such as P32: "I will work on two different projects, one using no generative AI, and using a lot of generative AI in the other one. I want to do this to test out how effective Gen AI is, how efficient it is, and how much I learn from using no AI to using AI."

\textbf{Understanding coding tools and fundamentals:}
Students set the goal to learn coding tools and fundamentals; for example, P19: "Learn more about coding and file fundamentals.", P31: "make sure i know my tools well before starting.", and P11: "In the future, I can do exactly what I said above: start with documentation, and increase the efficiency of my programming." Students also mentioned learning then making artifacts to help them, such as P17: "Have a list of resources I can use to solve errors or issues that I run into". This sometimes included using GenAI to aid learning, such as P25: "I am going to start using generative AI more for learning. I am going to ask more follow-up questions to better understand a particular topic..."

\textbf{Metacognitive and process goals:}
Students set goals related to metacognitive skills. Examples include P15: "Develop a consistent approach to using generative AI for debugging, understanding APIs, and optimizing code.", P10: "I won’t be afraid to use new tools to help myself, and I will know when to recognize what tools might be a hindrance to my learning.", and P23: "Learn how to fix mistakes in the coding process."

\textbf{Specific/Actionable Planning:}
Students set goals around the importance of planning, including specific and actionable goals, such as P15: "Allocate a fixed amount of time for planning and research at the start of every project.", 
P16: "Spend at least an hour planning before digging into writing code.",
and P6: "Set 30 minutes aside for me to plan before I start the actual session."

\textbf{Breaking down tasks:}
Students set the goal of breaking down larger tasks better. Examples include P6: "Break my work into chunks to limit the amount of time I might get distracted partway through a work session.", P11: "Write down my tasks somewhere and keep them up while I program.", and P29: "Work in chunks."

\textbf{Time Management:}
Students set goals related to time management strategies. Examples include P1: "Keep track of time spent on something to understand if it is wasting more time than needed.", P5: "Focus on timeliness in my future coding sessions.", and P23: "Code for longer periods of time in one sitting."

\section{Limitations}
Students did the reflection as part of a final assignment in the class due during finals. The timing of the assignment may also have reduced the depth of the student reflections, as they worked during a busy time. Students without submissions had final grades of 1 C grade, 2 B-, 1 B+, 1 A-. Excluding the reflection assignment points, students turning in the reflection had an average course grade of 87.7 versus not submitted had 83.9. The above excludes three students who already stopped attending and doing course assignments, specifically the last project milestone, before the reflection was assigned.

We limit our claims and results to the course and student population in our dataset. Given the novelty of our reflection design, it can likely be improved and potentially personalized for different student populations or course contexts. Different or more scaffolding might be required particularly for more novice computing students.

\section{Discussion}

We contribute a novel design for student reflection on their programming practices using comparative video analysis. Our approach also asks students to record themselves programming both with and without generative AI tools, then compare these sessions. Critically, we also used the DEAL framework \cite{ashGeneratingDeepeningDocumenting2009} to design scaffolding for reflection, providing students and guided prompts for describing, examining, and articulating learning from their experiences. Our results showed that this reflection assignment supported students' learning about generative AI, including its downsides, negative impacts, and limitations, while also supporting reflecting on more critical usage patterns and prompt engineering skills. Students also demonstrated learning about programming in general, such as the need for background understanding before coding, planning, human help-seeking, meta-cognitive insights about their learning processes, and even professional identity development. 

Our work demonstrates how video-based comparative reflection can broadly deepen learning in computing education, including for more critical GenAI use. Students noticed patterns and trade-offs in their behaviors, learned, and developed concrete goals for improvement. As computing education grapples with integrating GenAI \cite{pratherWideningGapBenefits2024a, margulieuxSelfRegulationSelfEfficacyFear2024}, scaffolded comparative reflection can help fill the wide gap identified in prior work \cite{pratherHypeComprehensiveReview2025} on how to help students critically reflect on then improve their GenAI use practices. This first version of our method even appears to provide a scalable way to help students develop more meta-cognitive awareness not only about their GenAI tool use, but also programming in general, and their learning processes, even without external faculty feedback (adding that would definitely help even more, and more research is needed). Completely unexpected, students even set goals to reflect more in the future. Regular reflective habits are key for lifelong learning throughout their careers as the tech landscape continues to evolve, especially given GenAI.

In our analysis we saw only a few or no students, when they did carefully scaffolded reflection, who appeared to be threatened by GenAI or give GenAI too much credence;  they thought about GenAI as something to use critically with pros and cons and to be used responsibly.
When scaffolded, the quality of the student reflection was often very high, and engaged with critical and ethical use of AI without being prompted to do so in the reflection questions themselves. When we properly scaffold and support their learning, at least in upper level classes, maybe we can engage with our students as partners in navigating generative AI. Future work and instructors should engage in exploring and refining scaffolded reflection in lower level courses, where results might be different or more scaffolding may be need to be designed.

Directly inspired by our results, we suggest instructors and researchers design and facilitate student self-experimentation by working on one project without GenAI and another with GenAI to test, as P32 said "how effective Gen AI is, how efficient it is, and how much I learn from using no AI [versus] using AI." Within that, and for future work on reflection and students acting on reflection longitudinally, we recommend follow-up with students with a survey and interviews to see how well they retain and act on their learning and goals they set from reflection. We also propose randomized quantitative evaluation of scaffolded comparative video reflection, such as during instead of just the end of a course. 

Inspired by our specific comparative video reflection design, we think there is a large and fruitful design space for student reflection and learning, based on adapting  qualitative research techniques. For example, having research participants review and reflect on videos is actually a qualitative research technique called video-stimulated recall \cite{zhaiSystematicReviewStimulated2024, geigerVideostimulatedRecallCatalyst2016}. Designers might simplify different qualitative research method practices with respect to replicability and rigor, for example, with new tools like a video reflection annotation tool \cite{leungAdoptionVideoAnnotation2021}. But the design space is much broader. Ponder your favorite research studies; our work was inspired by Prather et al's ``widening gap'' study on novice GenAI use \cite{pratherWideningGapBenefits2024a}.  Fundamentally, constructing knowledge happens in qualitative research, and students need to construct knowledge also.

Maybe we can even blur the line between reflective learning and research to, for example, more rapidly figure out what skills for good GenAI use are, then teach them \cite{pratherHypeComprehensiveReview2025}? What if students or  professionals using GenAI in industry recorded themselves while building more complex programming projects, then did scaffolded reflected on the skills and knowledge they used, and how that interacted with GenAI helping or hindering their progress and learning? Everyone might benefit. Researchers, industry professionals, instructors, and even students might do qualitative analysis of videos to help identify key skill building moments and define more useful, critical, and ethical GenAI-integrated curricula.

\balance
\bibliographystyle{acm}
\bibliography{sigcse} 

\end{document}